\documentclass[twocolumn]{IEEEtran} 

\usepackage[dvips]{graphicx}
\usepackage{amsmath,amssymb,amsfonts,threeparttable}

\def\BibTeX{{\rm B\kern-.05em{\sc i\kern-.025em b}\kern-.08em
    T\kern-.1667em\lower.7ex\hbox{E}\kern-.125emX}}

\allowdisplaybreaks

\newtheorem{theorem}{Theorem}

\newtheorem{lemma}{Lemma}

\begin{document}

\title{An Approximation of the Outage Probability for Multi-hop AF Fixed Gain Relay} 

\author{Jun Kyoung Lee, \emph{Student Member}, Janghoon Yang, \emph{Member}, and Dong Ku Kim,
\emph{Member}
\thanks{EDICS : CL.1.2.0, CL.1.2.1, CL.1.2.2}\thanks{
The authors are with the Department of Electrical and Electronic
Engineering, Yonsei University, Seoul, Korea. (email: {player72,
jhyang00, dkkim}@yonsei.ac.kr)}\thanks{Tel: +82-10-9530-5436,
Fax:+82-2-365-4504}}

\maketitle

\begin{abstract}
In this letter, we present a closed-form approximation of the outage
probability for the multi-hop amplify-and-forward (AF) relaying
systems with fixed gain in Rayleigh fading channel. The
approximation is derived from the outage event for each hop. The
simulation results show the tightness of the proposed approximation
in low and high signal-to-noise ratio (SNR) region.
\end{abstract}

\begin{keywords}
Wireless relay channel, multi-hop relay, outage probability
\end{keywords}

\section{Introduction}
\PARstart{O}{utage} probability is an important measure of system
performances in fading channel for reliability of link quality.
Although multihop AF fixed gain relays can be more preferable for
simple implementation, due to complication in analysis associated
with noise accumulation factor with multi-hop relay, there are not
many existing theoretical researches on them. In [1] and [2], the
outage probabilities of multihop decode-and-forward (DF) relay
systems were calculated and the optimum power allocation schemes
were proposed. In [1], it is argued that the outage probability of
multihop DF relay systems can be the lower bound of multihop AF
relay systems. In [3], Hasna et. al. found the outage probability
for multihop AF variable gain relay systems. In [4], Karagiannidis
provided the bounds of the outage probability for multihop AF fixed
gain relay systems by using harmonic and geometric mean, which are
not tight in high SNR region.

In this letter, we derive a closed-form approximation of the outage
probability for multihop AF fixed gain relay systems by a novel
approach considering the event space of the outage, which is tight
in all SNR regions. The remainder of this letter is organized as
follows. The system and channel model for the multihop AF fixed gain
relay system is presented, and the received SNR at the destination
is derived in Section II. In Section III, outage probability of
relaying system is derived for arbitrary number of hops. In Section
IV, we discuss the theoretical results and the simulation results
for the system. Finally, we conclude this letter in Section V.

\section{System model}
Considering the general $N$-hop relay network, there are $(N-1)$
relays between the source and the destination. It is assumed that
the relaying network is operated on time division multiplexing (TDM)
so that the transmission at any node occurs in different time slots
on the same carrier frequency.

Assuming that a signal at the source is transmitted with an average
power ${E_1 }$ and the fixed gain relays are serially placed from
the source to the destination, the instantaneous end-to-end SNR at
the destination can be written as
\begin{equation}
\begin{array}{l}
\gamma _N  = \frac{{A_{N - 1}^2 A_{N - 2}^2 \, \cdots A_1^2 \left|
{h_N } \right|^2 \left| {h_{N - 1} } \right|^2 \,\, \cdots
\,\,\left| {h_1 } \right|^2 E_1 }} {{\sum\limits_{j = 1}^{N - 1}
{\left( {\prod\limits_{i = j}^{N - 1} {A_i^2 \left| {h_{i + 1} }
\right|^2 } } \right)} \sigma _j^2  + \sigma _N^2 }}
\end{array}
\end{equation}
where $h_k $ is the fading amplitude of the channel at the $k$-th
hop with unit variance, i.e., $E\{ {\left| {h_k } \right|^2 } \} =
1$ where $k = 1,\,\, \cdots \,\,,\,\,N$, in which $E\{  \cdot  \}$
is the expectation operator, and $\sigma _k^2$ is the variance of
the additive white Gaussian noise (AWGN) with mean zero at the
$k$-th hop.

In (1), the amplification factor of the $l$-th relay with fixed gain
is defined [5] as
\begin{equation}
A_l  = \sqrt {\frac{{E_{l + 1} }} {{E_l  + \sigma _l^2 }}} \,\,\,\,,
\,\,l = 1,\,\, \cdots \,\,,\,\,N-1
\end{equation}

\section{Outage probability for multi-hop relaying systems with fixed gain}
The outage probability is defined as the probability that the
end-to-end SNR, $\gamma _N$ , falls below a threshold level of SNR.
For N-hop AF fixed gain relay system, the outage probability can be
written as
\begin{equation}
P_{out} (\gamma _N  < \gamma _{Th} )
\end{equation}
where $\gamma _{Th}$ is the threshold level of SNR.

Unfortunately, it is very difficult to obtain the closed-form
expression for the outage probability of multi-hop AF fixed gain
relay system due to the noise accumulation, as shown in the
denominator of (1). Alternatively, the outage probability of DF
relay in [1] or of AF variable gain relay in [3] were proposed to be
lower bounds for AF fixed gain relay. However, those can be
definitely lower bounds but are not tight. [4] found a lower bound
for AF fixed gain relay by using the well-known inequality between
harmonic and geometric mean. But it loses tightness in high SNR
region. In this letter, rather than finding a bound, we focus on
finding out a closed-form accurate approximation of the outage
probability for multihop AF fixed gain relay. To this end, the
following theorem is introduced first.

\begin{theorem} The outage probability for the N-hop AF relaying system with fixed gain is
lower bounded by $G_{1,N + 1}^{N,1} \left[ {\bar \gamma _{Th} \left|
{\begin{array}{*{20}c}
   1  \\
   {1,\,\,1,\,\, \cdots \,\,,\,\,1,0}  \\
\end{array}} \right.} \right]$, where $\bar \gamma _{Th}  = \frac{{\sigma _N^2 }}{{A_{N - 1}^2 A_{N - 2}^2 \, \cdots A_1^2 E_1 }}\gamma _{Th}
$.
\end{theorem}

\begin{proof} The end-to-end SNR of N-hop AF relay is upper bounded to $\tilde \gamma _N $ with the assumption that relays operate at asymptotically high SNR,
i.e., $\sigma _1^2  = \,\, \cdots \,\, = \sigma _{N - 1}^2  = 0$.
\begin{equation}
\gamma _N  \le \frac{{A_{N - 1}^2 A_{N - 2}^2 \, \cdots A_1^2 \left|
{h_N } \right|^2 \left| {h_{N - 1} } \right|^2 \,\, \cdots
\,\,\left| {h_1 } \right|^2 E_1 }}{{\sigma _N^2 }} \buildrel \Delta
\over = \tilde \gamma _N
\end{equation}
From (4), the outage probability is obviously lowered bounded as
\begin{equation}
P_{out} (\tilde \gamma _N  < \gamma _{Th} ) \le P_{out} (\gamma _N
< \gamma _{Th} )
\end{equation}
The left term of (5) can be equivalently expressed as
\begin{equation}
P_{out} (\gamma '_N  < \bar \gamma _{Th,N} ) \le P_{out} (\gamma _N
< \gamma _{Th} )
\end{equation}
where $\gamma '_N  = \left| {h_1 } \right|^2 \left| {h_2 } \right|^2
 \cdots \left| {h_N } \right|^2 $ and $\bar \gamma _{Th,N} =
\frac{{\sigma _N^2 }}{{A_{N - 1}^2 A_{N - 2}^2 \, \cdots A_1^2 E_1
}}\gamma _{Th} $.

By using Weibull distribution which is the general form of the
exponential distribution family, the PDF of the cascaded exponential
random variables, $\gamma =\left| {h_1 } \right|^2 \left| {h_2 }
\right|^2 \,\, \cdots \,\,\left| {h_N } \right|^2$, can be expressed
by using [6, eq. (3)] as
\begin{equation}
f_{\gamma}  (\gamma ) = \frac{1} {\gamma }G_{N,0}^{0,N} \left[
{\frac{1} {\gamma }} \bigg\vert{ {\begin{array}{*{20}c}{0,\,\,0,\,\,
\cdots \,\,,\,\,0}
\\
{-} \end{array}} } \right]
\end{equation}
where $G\left[  \cdot  \right]$ is the Meijer G function in [7, eq.
(9.301)] which is a built-in function in the well-known softwares
such as MAPLE and MATHEMATICA.

\begin{figure}[!t]
   \centering
    \includegraphics[width=3.4in]{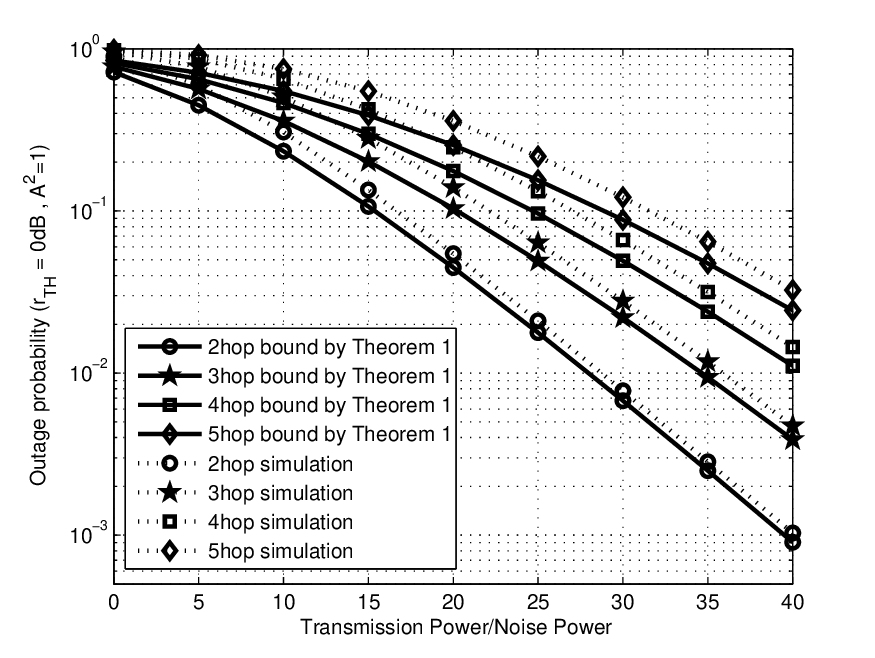}
    \caption{The bound in Theorem 1 of outage probability for multihop AF relay system with fixed gain ($A_1^2  =  \cdots  = A_{N - 1}^2  = 1$).}
    \label{Fig 3}
\end{figure}

The lower bound of the outage probability for the N-hop AF fixed
gain relay can be calculated with the help of [8, eq.
(07.34.16.0002.01)] and [8, eq. (07.34.21.0001.01)] as
\begin{equation}
\begin{array}{l}
P_{out} (\tilde \gamma _N  < \gamma _{Th} ) = P_{out} (\gamma '_N  <
\bar \gamma _{Th,N})\\
= \int\limits_0^{\bar \gamma _{Th,N} } {\frac{1}{\gamma
}G_{N,0}^{0,N} \left[ {\left. {\frac{1}{\gamma }}
\right|\begin{array}{*{20}c}
   {0,\,\,0,\,\, \cdots \,\,,\,\,0}  \\
    -   \\
\end{array}} \right]} \,d\gamma  \\
= G_{1,N + 1}^{N,1} \left[ {\bar \gamma _{Th,N} \left|
{\begin{array}{*{20}c}
   1  \\
   {1,\,\,1,\,\, \cdots \,\,,\,\,1,0}  \\
\end{array}} \right.} \right] \\
 \end{array}
\end{equation}
\end{proof}

As shown in Fig. 1, Theorem 1 is confirmed as a lower bound of
outage probability for multihop AF fixed gain relay by the
simulation, assuming that all amplification factors of the relays
are 1. However, (8) loses tightness in low SNR region, since the
closed-form lower bound is derived under the high SNR assumption.
Therefore, the different method should be considered to increase
accuracy in low SNR region.

\begin{lemma} In N-hop AF fixed gain relay network, as the number of hops is increased,
the end-to-end SNR is decreased, $\gamma _N  \le \gamma _{N - 1} \le
\,\, \cdots \,\, \le \gamma _2  \le \gamma _1 $, while outage
probability is increased, $P_{out} (\gamma _1  < \gamma _{Th} ) \le
P_{out} (\gamma _2  < \gamma _{Th} ) \le  \cdots  \le P_{out}
(\gamma _N  < \gamma _{Th} )$.
\end{lemma}

\begin{proof} It can be easily proved.
\end{proof}

\begin{figure}[!t]
   \centering
    \includegraphics[width=3.2in]{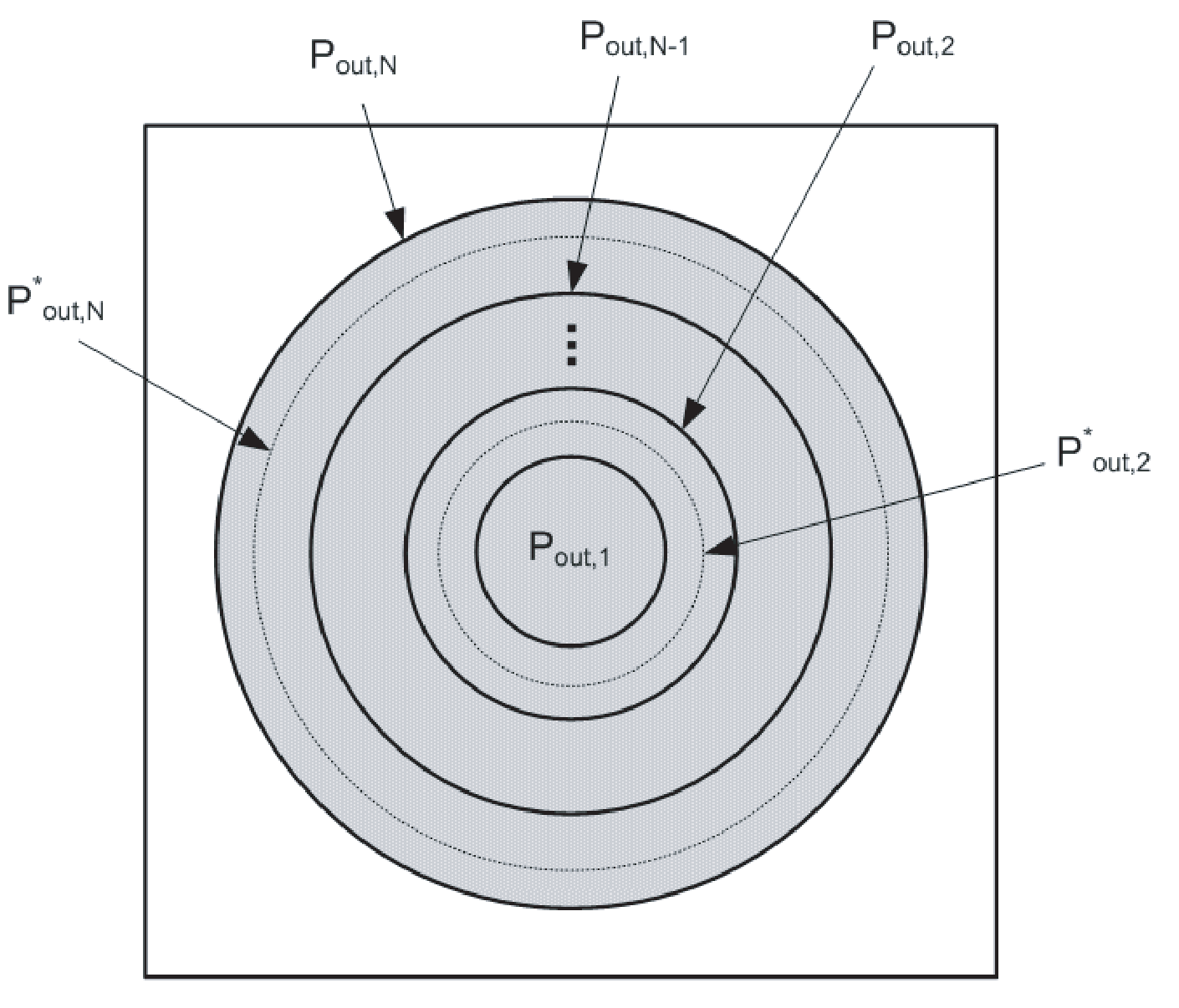}
    \caption{The outage event space for multihop AF relaying system with fixed gain.}
    \label{Fig 3}
\end{figure}

For notational simplicity, let $P_{out,n}$ and $P_{out,n}^*$ for $n
= 1,2, \cdots N$ be denoted by $P_{out} (\gamma _n < \gamma _{Th} )$
and $P_{out} (\tilde \gamma _n  < \gamma _{Th} )$, respectively.
From Theorem 1 and Lemma 1, the outage event space for multihop AF
relaying system with fixed gain can be drawn as Fig. 2.

As shown in Fig. 2, the outage probability can be expressed as the
sum of the probabilities for each hop. The outage probability can be
evaluated for each number of hops in the following way.

(1) 1-hop case:
\begin{equation}
P_{out,1}  = P(\gamma _1  < \gamma _{Th} ) = P_{out} (\gamma '_1  <
\bar \gamma _{Th,1} ) = 1 - e^{ - \bar \gamma _{Th,1} }
\end{equation}

(2) 2-hop case:
\begin{equation}
\begin{array}{l}
 P_{out,2}  = P(\gamma _2  < \gamma _{Th} ) \\
 = P(\gamma _1  < \gamma _{Th} ,\;\gamma _2  < \gamma _{Th} ) + P(\gamma _1  > \gamma _{Th} ,\;\gamma _2  < \gamma _{Th} ) \\
 \end{array}
\end{equation}
where the outage probability cannot be directly calculated since
$\gamma _1$ and $\gamma _2$ are not independent. However, it can be
approximated with the help of Fig. 2. The first term in (10) is
obviously rewritten as $P(\gamma _1  < \gamma _{Th} )$ and the
second term is approximately calculated as $P(\gamma _1  > \gamma
_{Th} )P(\gamma _2  < \gamma _{Th} )$ for simplicity, as if the two
events are independent. Since it is hard to have a closed form
expression for $P(\gamma _2  < \gamma _{Th} )$, it can approximated
by using the lower bound in Lemma 1.

Therefore, the approximation of the outage probability for the 2-hop
case can be obtained as

\begin{equation}
\begin{array}{l}
 P_{out,2}  \approx P(\gamma '_1  < \bar \gamma _{Th,1} ) + P(\gamma '_1  > \bar \gamma _{Th,1} )P(\gamma _2  < \gamma _{Th} ) \\
 \,\,\,\,\,\,\,\,\,\,\,\,\,\,\,\,\, \ge P_{out,1}  + P(\gamma '_1  > \bar \gamma _{Th,1} )P(\gamma '_2  < \bar \gamma _{Th,2} ) \buildrel \Delta \over = \tilde P_{out,2}  \\
 \end{array}
\end{equation}

(3) 3-hop case:
\begin{equation}
\begin{array}{l}
 P_{out,3}  = P(\gamma _3  < \gamma _{Th} ) \\
 = P(\gamma _1  < \gamma _{Th} ,\;\gamma _2  < \gamma _{Th} ,\gamma _3  < \gamma _{Th} ) \\
 \,\,\,\,\,+ P(\gamma _1  > \gamma _{Th} ,\;\gamma _2  < \gamma _{Th} ,\gamma _3  < \gamma _{Th} ) \\
 \,\,\,\,\,+ P(\gamma _1  > \gamma _{Th} ,\;\gamma _2  > \gamma _{Th} ,\gamma _3  < \gamma _{Th} ) \\
 \approx \tilde P_{out,2} \, + P(\gamma _3  < \gamma _{Th} )P(\gamma _1  > \gamma _{Th} ,\;\gamma _2  > \gamma _{Th} ) \\
 \approx \tilde P_{out,2}  + P(\gamma '_3  < \bar \gamma _{Th,3} )P(\gamma _3  > \gamma _{Th} ) \\
 = \tilde P_{out,2}  + P(\gamma '_3  < \bar \gamma _{Th,3} )(1 - P(\gamma _3  < \gamma _{Th} )) \\
 \approx \tilde P_{out,2}  + P(\gamma '_3  < \bar \gamma _{Th,3} )(1 - P(\gamma '_3  < \bar \gamma _{Th,3} )) \buildrel \Delta \over = \tilde P_{out,3}  \\
 \end{array}
\end{equation}

Similarly, for general N-hop AF relaying system with fixed gain, its
outage probability can be approximated as

(4) N-hop case:
\begin{equation}
\begin{array}{l}
 P_{out,N}  = P(\gamma _N  < \gamma _{Th} ) \\
 \approx \tilde P_{out,N - 1}  + P(\gamma '_N  < \bar \gamma _{Th,N} )(1 - P(\gamma '_N  < \bar \gamma _{Th,N} )) \\
 \,\,\,\,\,\,\, \buildrel \Delta \over = \tilde P_{out,N}  \\
 \end{array}
\end{equation}

To increase the accuracy of the approximation, we perform averaging
out the noise power in (1) and applying it to the right term in
(13). If $\sigma _1^2  = \,\, \cdots \,\, = \sigma _{N }^2$, then
the approximation of the outage probability can be written as
\begin{equation}
\begin{array}{l}
P_{out,N}  \approx \tilde P_{out,N - 1}  + P\left( {\gamma '_N  <
\bar \gamma _{Th,N} } \right)\,\,\,\,\,\,\,\,\,\,\,\,\,\,\,\,\,\,\,\,\,\,\,\,\,\,\,\,\\
\,\,\,\,\,\,\,\,\,\,\,\,\,\,\,\,\,\,\,\,\, \cdot \left( {1 - P\left(
{\gamma '_N < \left[ {\sum\limits_{j = 1}^{N - 1} {\left(
{\prod\limits_{i = j}^{N - 1} {A_i^2 } } \right)} + 1} \right]\bar
\gamma _{Th,N} } \right)} \right)
\end{array}
\end{equation}

Without loss of generality, for $N \ge 3$, the outage probability in
(14) can be rewritten as
\begin{equation}
\begin{array}{l}
 P_{out,N}  \\
  \approx \left( {1 - e^{ - \bar \gamma _{Th,1} } } \right) + e^{ - \bar \gamma _{Th,1} } G_{1,3}^{2,1} \left[ {\bar \gamma _{Th,2} \left| {\begin{array}{*{20}c}
   1  \\
   {1,\,\,1,\,\,0}  \\
\end{array}} \right.} \right] \\
+ \sum\limits_{n = 3}^N {G_{1,n + 1}^{n,1} \left[ {\bar \gamma
_{Th,n} \left| {\begin{array}{*{20}c}
   1  \\
   {1,\,\,1,\,\, \cdots \,\,,\,\,1,0}  \\
\end{array}} \right.} \right]} \left( {1 - \begin{array}{*{20}c}
   {}  \\
   {}  \\
\end{array}} \right. \\
 \left. {G_{1,n + 1}^{n,1} \left[ {\left( {\sum\limits_{j = 1}^{n - 1} {\left( {\prod\limits_{i = j}^{n - 1} {A_i^2 } } \right)}  + 1} \right)\bar \gamma _{Th,n} \left| {\begin{array}{*{20}c}
   1  \\
   {1,1, \cdots ,1,0}  \\
\end{array}} \right.} \right]} \right) \\
 \end{array}
\end{equation}

\section{Simulation results}
The proposed approximation and the simulation results of the outage
probability for multihop AF relaying system with fixed gain are
shown in Fig. 3. For the simulation, the threshold level of SNR is
0dB. From the results, it is noted that the proposed approximation
of the outage probability is very close to the simulation result in
low and high SNR region.

\begin{figure}[!t]
   \centering
    \includegraphics[width=3.6in]{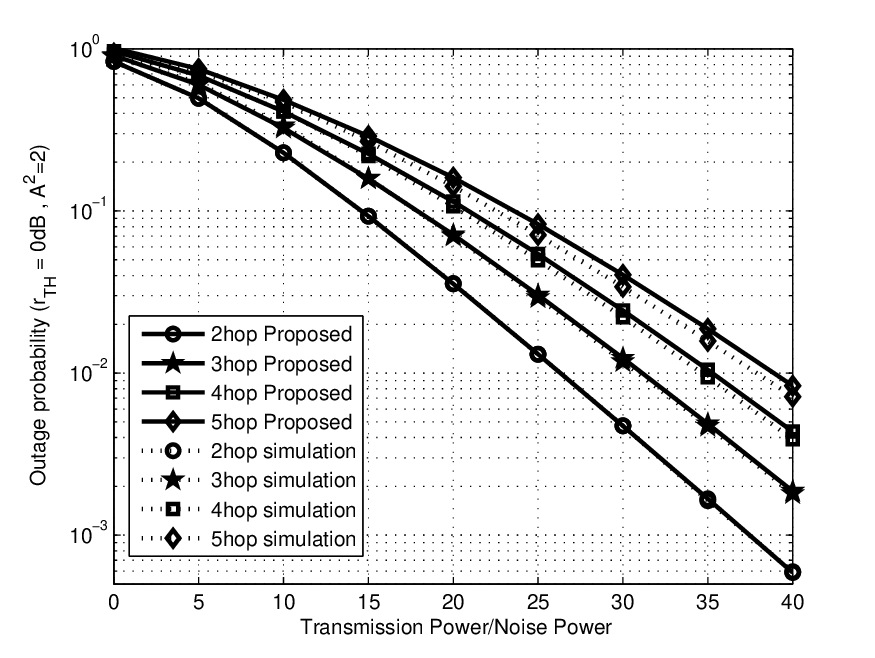}
    \caption{The proposed approximation of outage probability for multihop AF relay system with fixed gain ($A_1^2  =  \cdots  = A_{N - 1}^2  = 2$).}
    \label{Fig 3}
\end{figure}

\section{Conclusion}
In this letter, an accurate approximation of the outage probability
for the multihop AF relaying systems with fixed gain in Rayleigh
fading channel is derived by using the outage event space. The
numerical results show that it provides the very accurate
approximation in all SNR region.

\nocite{*}
\bibliographystyle{IEEE}

\end{document}